\documentclass[12pt,preprint]{aastex}
\shorttitle{Cosmic-Ray Modified Shocks}
\shortauthors{Kang {\it et al.~}}
\def\etal{{\it et al.}}
\def\eg{{\it e.g.,~}}
\def\ie{{\it i.e.,~}}

\def\kms{~{\rm km~s^{-1}}}
\def\cm3{~{\rm cm^{-3}}}

\begin{document}
\title{Diffusive Shock Acceleration in Test-Particle Regime}

\author{Hyesung Kang\altaffilmark{1}
        and Dongsu Ryu\altaffilmark{2}}

\altaffiltext{1}
{Department of Earth Sciences, Pusan National University, Pusan 609-735,
Korea:\\ kang@uju.es.pusan.ac.kr}
\altaffiltext{2}
{Department of Astronomy and Space Science, Chungnam National University,
Daejeon 305-764, Korea:\\ ryu@canopus.cnu.ac.kr}

\begin{abstract}
We examine the test-particle solution for diffusive shock acceleration,
based on simple models for thermal leakage injection and Alfv\'enic drift.
The critical injection rate, $\xi_c$, above which the cosmic ray (CR) pressure
becomes dynamically significant, depends mainly on the sonic shock Mach number,
$M$, and preshock gas temperature, $T_1$.  
In the hot-phase interstellar medium (ISM) and intracluster medium,
$\xi_c \la 10^{-3}$ for shocks with $M \la 5$, 
while $\xi_c \approx 10^{-4}(T_1/10^6{\rm K})^{1/2}$ for shocks with $M\ga 10$.
For $T_1=10^6$ K, for example, the test-particle solution would be valid 
if the injection momentum, $p_{\rm inj} > 3.8 p_{\rm th}$ 
(where $p_{\rm th}$ is thermal momentum).
This leads to the postshock CR pressure less than 10\% of the shock ram pressure.
If the Alfv\'en speed is comparable to the sound speed in the preshock flow,
as in the hot-phase ISM, the power-law slope of CR spectrum can be significantly 
softer than the canonical test-particle slope.
Then the CR spectrum at the shock can be approximated by the {\it revised}
test-particle power-law with an exponential cutoff at the highest 
accelerated momentum, $p_{\rm max}(t)$.
An analytic form of the exponential cutoff is also suggested.

\end{abstract}


\section{Introduction}

Suprathermal particles are produced as an inevitable consequence of
the formation of collisionless shocks in tenuous astrophysical plasmas 
and they can be further accelerated to very high
energies through interactions with resonantly scattering Alfv\'en 
waves in the converging flow across a shock \citep{bell78, dru83,
blaeic87,maldru01}. 
The most attractive feature of the diffusive shock acceleration (DSA) 
theory is the simple prediction of the power-law momentum distribution 
of cosmic rays (CRs), $f(p) \propto p^{-3\sigma/(\sigma-1)}$ 
(where $\sigma$ is the shock compression ratio) in the test particle regime. 
For strong, adiabatic gas shocks, this gives a power-law
index of 4, which is reasonably close to the observed, `universal' index
of the CR spectra in many environments.

The nonthermal particle injection and ensuing acceleration at shocks 
depend mainly upon the shock Mach number, field obliquity angle, 
and the strength of the Alfv\'en turbulence responsible for scattering.
At quasi-parallel shocks, the shock Mach number is the primary parameter 
that determines the CR acceleration efficiency,
while the injection fraction, $\xi$ (the ratio of CR particles to the total
particles passed through the shock), is the secondary parameter.
Detailed nonlinear treatments of DSA predict that at strong shocks,
with a small fraction of $\xi > 10^{-4}$, 
a significant fraction of the shock kinetic energy is transferred 
to CRs and there are highly nonlinear back-reactions from CRs 
to the underlying flow \citep{berezhko97, kj07}.
Indeed, multi-band observations of nonthermal radio to $\gamma$-ray 
emissions from several supernova remnants (SNRs) 
have been successfully explained by efficient DSA features such as 
high degree of shock compression and amplification of magnetic fields in
the precursor \citep[e.g.][]{reynolds08,berezhko09,morlino09}.

It has been recognized, however, that the CR spectrum at sources, $N(E)$,
predicted for shocks strongly modified by CR feedback may be too flat to be consistent
with the observed flux of CR nuclei at Earth, $J(E)$. 
Recently \citet{ave09} analyzed the spectrum of CR nuclei up 
to $\sim 10^{14}$ eV measured by TRACER instruments and found that 
the CR spectra at Earth can be fitted by a single power law of 
$J(E) \propto E^{-2.67}$.
Assuming an energy-dependent propagation path length 
($\Lambda \propto E^{-0.6}$), they
suggested that a soft source spectrum, $N(E)\propto E^{-s}$ 
with $s \sim 2.3-2.4$, is preferred by the observed data.
This is much softer than the CR spectrum that the nonlinear DSA predicts for
strong SNRs, which are believed to be the main
accelerators for Galactic CRs up to the knee energy around $10^{15.5}$eV.
Thus, in order to reconcile the DSA prediction with the observed $J(E)$,
the bulk of Galactic CRs should originate from
SNRs in which the CR acceleration efficiency is 10 \% or so
(\ie roughly in the test-particle regime).
Such inefficient acceleration could be possible for SNRs in the hot
phase of the interstellar medium (ISM) (\ie low shock Mach number shocks) and
for the inject fraction smaller than $10^{-4}$ \citep{kang10}.

The scattering by Alfv\'en waves tends to isotropize
the CR distribution in the wave frame, which may drift upstream 
at Alfv\'en speed with respect to the bulk plasma \citep{skill75}.
This Alfv\'enic drift in the upstream region reduces the velocity jump
that the particles experience across the shock, which in turn softens 
the CR spectrum beyond the canonical test-particle slope 
($s=2$ for strong shocks) \citep{kang10, capri10}. 
Moreover, the Alfv\'enic drift in {\it amplified 
magnetic fields both upstream and downstream} can drastically soften the 
accelerated particle spectrum even in nonlinear modified shocks
\citep{zp08,pzs10}.

At collisionless shocks suprathermal particles moving faster than the
postshock thermal distribution may swim through the MHD waves and 
leak upstream across the shocks and get injected into the CR population
\citep{mal98,gies00, kjg02}.
But it is not yet possible to make precise quantitative predictions for
the injection process from first principles,
because complex plasma interactions among CRs, waves, and the underlying 
gas flow are not fully understood yet \citep[\eg][]{maldru01}.
Until plasma simulations such as hybrid or particle-in-cell simulations
reach the stage where the full problem can be treated 
with practical computational resources,
in the studies of DSA we have to adopt a phenomenological injection scheme
that can emulate the injection process. 

In this paper, we will examine the relation between the thermal
leakage injection model described in \citet{kjg02} and the time-dependent
test-particle solutions for DSA.
The basic models are described in \S 2,
while the analytic expression for the CR spectrum in the test-particle limit 
is suggested in \S 3.
Finally, a brief summary will be given in \S 4.

\section{BASIC MODELS}

%the basics of DSA
In the kinetic DSA approach,
the following diffusion-convection equation 
for the pitch-angle-averaged distribution function, $f(x,p,t)$, 
is solved along with suitably modified gasdynamic equations:
\begin{equation}
\label{diffcon}
{\partial f \over \partial t}  + (u+u_w) {\partial f \over \partial x}
= {p \over 3} {{\partial (u+u_w)} \over {\partial x}} 
{{\partial f} \over {\partial p}} 
+ {\partial \over \partial x} \left[\kappa(x,p)
{\partial f \over \partial x} \right],
\end{equation}
where $\kappa(x,p)$ is the spatial diffusion coefficient
and $u_w$ is the drift speed of the local Alfv\'enic wave turbulence 
with respect to the plasma \citep{skill75}.
We consider only the proton CR component.

\subsection{Alfv\'enic Drift Effect}

Since the Alfv\'en waves upstream of the subshock are expected to be established
by the streaming instability, the wave speed is set there to be $u_w=-v_A$.
Downstream, it is likely that the Alfv\'enic turbulence is nearly
isotropic, hence $u_w=0$ there.
As a result, the velocity jump across the shock is reduced, and
the slope of test-particle plower-law spectrum should be revised as
\begin{equation}
q_{\rm tp}={{3(u_1-v_A)}\over u_1-v_A-u_2}
={{3\sigma(1-M_A^{-1})}\over (\sigma-1-\sigma M_A^{-1})},
\label{qtp}
\end{equation}
where $u_1$ and $u_2$ are the upstream and downstream speed, respectively,
in the shock rest frame, 
$\sigma = u_1/u_2=\rho_2/\rho_1$ is the shock compression ratio,
and $v_A$ and $M_A=u_1/v_A$ are the Alfv\'en speed 
upstream and Alfv\'en Mach number.
Hereafter, we use the subscripts '1', and '2' to denote
conditions upstream and downstream of the shock, respectively.
Thus the CR spectrum would be softer than the canonical power-law spectrum 
with the slope, $3\sigma/(\sigma-1)$, unless $M_A\gg 1$.

The left panel of Figure 1 shows the {\it revised} test-particle slope 
$q_{\rm tp}$ as a function of the sonic Mach number, $M$, 
for different Alfv\'en speeds, $v_A= \delta \cdot c_s$
(where $c_s$ is the upstream sound speed).
In the hot-phase ISM of $T \approx 10^6$K 
with the hydrogen number density $n_H \approx 0.003~ {\rm cm}^{-3}$ and
the magnetic field strength $B \approx 5\mu$G, 
the sound speed is $c_s \approx 150 \kms$ and
the Alfv\'en speed is $v_A \approx 170 \kms$.  
So $\delta \approx 1$ is a representative value.
If $\delta \approx (P_B/P_g)^{1/2} \approx 1$,
the Alfv\'en drift effect is significant
for Alfv\'en Mach number, $M_A\approx M \la  30$.
Consequently, this effect reduces the CR acceleration efficiency. 
Of course, it is not important for strong
shocks with $u_s \gg c_s \sim v_A$ (\ie $M_A \ga 30$).

\subsection{Thermal Leakage Injection Model}

Since the velocity distribution of suprathermal particles is not
isotropic in the shock frame, the diffusion-convection equation cannot directly
follow the injection from the non-diffusive thermal pool into the diffusive 
CR population.
Here we adopt the thermal leakage injection model that was originally
formulated by \citet{gies00} based on the calculations of \citet{mal98}. 
In this model particles above a certain injection momentum $p_{\rm inj}$ 
cross the shock and get injected to the CR population.
We adopt a smooth ``transparency function'', 
$\tau_{\rm esc}(\epsilon_B, v)$, that expresses the probability of 
suprathermal particles at a given velocity, $v$, 
leaking upstream through the postshock MHD waves.
One free parameter controls this function;
$\epsilon_B = B_0/B_{\perp}$, the ratio of
the general magnetic field along the shock normal, $B_0$, to
the amplitude of the postshock MHD wave turbulence, $B_{\perp}$. 
Although plasma hybrid simulations and theories both suggested that
$0.25 \la \epsilon_B \la 0.35$ \citep{mv98}, the physical range of
this parameter remains to be rather uncertain due to lack of
full understanding of relevant plasma interactions.
Since $\tau_{\rm esc}$ increases gradually from zero to one in
the thermal tail distribution, the ``effective'' injection momentum 
can be approximated by
\begin{equation}
p_{\rm inj} \approx 1.17 m_p u_2 (1+ {1.07 \over \epsilon_B})
\equiv Q_{\rm inj}(M, \epsilon_B) p_{\rm th}
\label{pinj}
\end{equation}
where $p_{\rm th}= \sqrt{2m_p k_B T_2}$ is the thermal peak momentum of
the immediate postshock gas with temperature $T_2$ 
and $k_B$ is the Boltzmann constant \citep{kjg02}.

The right panel of Figure 1 shows the value of $Q_{\rm inj}$ as a function of
$M$ for three values of $\epsilon_B= 0.21$, 0.23, and 0.27,
which represents ``inefficient'', ``moderately efficient'', ``efficient'' injection cases,
respectively (see Fig. 4 below).
At weaker shocks the compression is smaller and so the ratio $u_2/u_1$ is larger. 
For stronger turbulence (larger $B_{\perp}$, smaller $\epsilon_B$) 
it is harder for particles to swim across the shock. 
So for both of these cases, $p_{\rm inj}$ has to be larger.
Hence the value of $Q_{\rm inj}(M, \epsilon_B)$ is larger
for weaker shocks and for smaller $\epsilon_B$,
which leads to a lower injection fraction.

In our thermal leakage injection model, 
the CR distribution function at $p_{\rm inj}$ is then anchored to the 
postshock Maxwellian distribution as,
\begin{equation}
f_{\rm inj}= f(p_{\rm inj})= {n_2 \over \pi^{1.5}}~ p_{\rm th}^{-3} ~\exp(-Q_{\rm inj}^2),
\label{finj}
\end{equation}
where $n_2$ is the postshock proton number density and 
the distribution function is defined in general as 
$\int 4\pi p^2f(p) dp = n$.
For the test-particle power-law spectrum,
the value of $Q_{\rm inj}$ determines the amplitude of the subsequent 
suprathermal power-law distribution as 
$f(p)=f_{\rm inj}\cdot (p/p_{\rm inj})^{-q_{\rm tp}}$.
Then the CR injection fraction can be defined as 
\begin{equation}
\xi \equiv {n_{CR} \over n_2 } = {4 \over \sqrt{\pi}} Q_{\rm inj}^3 
\exp(-Q_{\rm inj}^2) {1 \over {q_{\rm tp}(M) - 3}},
\end{equation}
which depends only on the ratio $Q_{\rm inj}$ and the slope $q_{\rm tp}$, 
but not on the postshock temperature $T_2$.
For $Q_{\rm inj}=3.8$, for example, $\xi = 6.6 \times 10^{-5}/(q_{\rm tp} - 3)$,
which becomes  $\xi = 6.6 \times 10^{-5}$ for strong shocks with 
$q_{\rm tp}=4.0$.

\subsection{Bohm-type Diffusion Model}
In modeling DSA, it is commonly assumed that
the particles are resonantly scattered by self-generated waves,
so the Bohm diffusion model can represent a saturated
wave spectrum (\ie the mean scattering length, $\lambda = r_g$, where $r_g$ is the gyro-radius).
Here, we adopt a Bohm-type diffusion coefficient
that includes a weaker non-relativistic momentum dependence,
\begin{equation}
\kappa(x,p) = \kappa^* \cdot ({p \over {m_p c}})^{\alpha} \left [{\rho(x) \over \rho_1}
\right]^{-m},
\label{kappa}
\end{equation}
where the coefficient $\kappa^*= m_p c^3/(3eB_0)$ depends on the upstream mean field strength.
The case with $m=1$ approximately accounts for the compressive amplification of Alfv\'en waves. 

The mean acceleration time for a particle to reach $p_{\rm max}$ from $p_{\rm inj}$
in the test-particle limit of DSA theory can be approximated  by 
\begin{equation}
t_{\rm acc} = {3\over {u_1-v_A- u_2}} \int_{p_{\rm inj}}^{p_{\rm max}}
\left({\kappa_1\over {u_1-v_A}} + {\kappa_2\over u_2} \right) {dp \over p},
\label{tacc}
\end{equation}
if we assume the bulk drift of waves with $v_A$ in the upstream region \cite[\eg][]{dru83}.
Then the maximum momentum can be estimated by setting $t = t_{\rm acc}$ as
\begin{equation}
p_{\rm max}(t)^{\alpha} \approx {{\alpha (1-M_A^{-1})(\sigma -1 -\sigma M_A^{-1})}
\over {3\sigma[1+(1-M_A^{-1})\sigma^{1-m}]}}
{u_s^2 \over \kappa^*} t = f_c{u_s^2 \over \kappa^*} t ,
\label{pmax}
\end{equation}
where $u_s=u_1$ is the shock speed \citep{krj09}.
For the case of $m=1$, the typical value of the parameter, $f_c= 
\alpha (1-M_A^{-1})(\sigma -1 -\sigma M_A^{-1})/
\{3\sigma[1+(1-M_A^{-1})\sigma^{1-m}]\}$, is $\sim 1/8$ in the limit of 
$M_A\gg 1$ and $M\gg 1$. 

\section{TEST-PARTICLE SPECTRUM}

If the injection is inefficient, especially at weak shocks,
the CR pressure remains dynamically insignificant and the test-particle
solution is valid.
Caprioli \etal (2009) (CBA09 hereafter) derived the analytic solution for a {\it steady-state}, 
test-particle shock with a free-escape boundary (FEB) at a distance $x_{\rm FEB}$ upstream of
the shock (\ie $f(x>x_{\rm FEB})=0$). 
For a diffusion coefficient that depends on the momentum as
$\kappa(p)= \kappa^* (p/m_p c)^{\alpha}$, the CR distribution at the shock location, $x_s$,
is given by 
\begin{equation}
f_{\rm tp}(x_s,p) = f_0 \cdot \exp \left[- q_{\rm tp}
 \int_{z_{\rm inj}}^z {dz' \over z'} {1\over {1-\exp(-1/z'^{\alpha})} }\right],
\label{fCBA}
\end{equation}
where $z=p/p^*$, $z_{\rm inj} = p_{\rm inj}/p^*$, $f_0=f_{\rm inj}$, 
and $p^*/m_p c= (x_{\rm FEB}u_s/\kappa^*)^{1/\alpha}$ is the cutoff momentum set by the FEB.
This expression can be re-written as,
\begin{equation}
f_{\rm tp}(x_s, p) = f_{\rm inj} \cdot ({p \over p_{\rm inj} })^{-q_{\rm tp}} \cdot 
\exp\left[- q_{\rm tp} C(z) \right],
\label{ftest}
\end{equation}
where the function $C(z)$ is given by 
\begin{equation}
C(z) = \int_{z_{\rm inj}}^z {dz' \over z'} {1\over {\exp(1/z'^{\alpha})-1} }.
\label{cofz}
\end{equation}
We show the function $C(z)$ for $\alpha=0.5$ and 1 in the left panel of Figure 2.
For $z\ll1$, $C(z)$ is small and so $\exp\left[- q_{\rm tp} C(z) \right]=1$, 
as expected.
For $z\gg1$, $C(z) \approx  z^{\alpha}=(p/p^*)^{\alpha}$. 
But this regime ($p\gg p^*$) is not really relevant, because the resulting
$f_{\rm tp}(x_s, p)$ is extremely small.
We are more interested in the exponential cutoff where $p\sim p^*$. 
Figure 2 shows that $C(z)$ increases much faster than $z^{\alpha}$ near $z\sim1$. 
In fact, at $z\sim 1$, approximately $C(z)\approx 0.29 z^2$ for $\alpha=1$
and $C(z) \approx 0.58 z$ for $\alpha=1/2$.
Thus equation (\ref{ftest}) can be approximated by 
\begin{equation}
f_{\rm tp}(x_s, p) \approx f_{\rm inj} \cdot ({p \over p_{\rm inj} })^{-q_{\rm tp}} \cdot 
\exp\left[- {0.29 q_{\rm tp} \over \alpha} ({p \over p^*})^{2\alpha} \right].
\label{ftest2}
\end{equation}

\citet{krj09} showed that the shock structure and the CR spectrum of time-dependent, CR modified
shocks with ever increasing $p_{\rm max}(t)$ are similar to those of steady-state shocks 
with particles escaping through the upper momentum boundary,
\ie $f(p>p_{\rm ub})=0$, if compared when $p_{\rm max}(t)=p_{\rm ub}$ (see their Figs. 10-11). 
They also showed that the exponential cutoff in the form of $\exp[-k(p/p_{\rm max})^{2\alpha}]$ 
matches well the DSA simulation results for CR modified shocks.
In the same spirit, we suggest that equation (\ref{ftest}) could represent the CR spectrum
at the shock location for time-dependent, test-particle shocks without particle escape,
in which the cutoff momentum is determined by the shock age as in equation (\ref{pmax}), 
\ie $p^* \sim p_{\rm max}(t)$. 

The distribution function $f(x,p_{\rm max})$ in the upstream region decreases
roughly as $\exp[-x/l_d(p_{\rm max})]$, where 
the diffusion length for $p_{\rm max}$ is 
\begin{equation}
l_d(p_{\rm max})= { \kappa(p_{\rm max}) \over u_s} = f_c u_s t.
\end{equation}
CBA09 spectrum in equation (\ref{ftest}) was derived from the FEB 
condition of $f(x>x_{\rm FEB},p)=0$ for steady-state shocks, 
while $f(x,p)\rightarrow 0$ only at $x\rightarrow \infty$ (upstream infinity)
for time-evolving shocks without particle escape.
So we presume that the cutoff momentum can be found by setting 
the location of FEB at $x_{\rm FEB} = \zeta \cdot l_d(p_{\rm max})$, where $\zeta \sim 1$.
From the condition that $p^*/m_p c= (\zeta l_d(p_{\rm max}) u_s/\kappa^*)^{1/\alpha}$,
we find $p^*= \zeta \cdot p_{\rm max}$.  

The right panel of Figure 2 shows the test-particle solution from
a time-dependent DSA simulation, in which the dynamical feedback of the CR pressure
was turned off.
Contrary to CBA09 case, no FEB is enforced in this simulation, so the
shock does not approach to a steady state, but instead evolves in time.
As the CRs are accelerated to ever high energies ($p_{\rm max} \propto t$), 
the scale length of the CR pressure increases linearly with time, 
$l_d(p_{\rm max}) \propto u_s t$.
So the shock structure evolves in a self-similar fashion, depending 
only on the similarity variable, $x/(u_s t)$ 
\citep[see][]{kj07}.
By setting $p^*= 1.2p_{\rm max}(t)$ (\ie $\zeta=1.2$)
and also by adopting the value of $f_{\rm inj}$ from the DSA simulation result, 
we calculated $f_{\rm tp}(x_s,p)$ according to equation (\ref{ftest}).  
As can be seen in the figure, the agreement between the numerical DSA results
and the analytic approximation is excellent.
Thus we take equation (\ref{ftest}) as the test-particle spectrum from DSA,
where $q_{\rm tp}$, $p_{\rm inj}$, $f_{\rm inj}$, and 
$p^* \approx 1.2p_{\rm max}(t)$ are given by 
equations (\ref{qtp}), (\ref{pinj}), (\ref{finj}), and (\ref{pmax}), respectively.

Figure 3 shows some examples of the test-particle spectrum given in equation (\ref{ftest}). 
We consider the shocks propagating
into the hot-phase of the ISM of $T_1=10^6$K or a typical intracluster
medium (ICM) of $T_1=10^7$K.  
The shock speed is given by $u_s=M \cdot c_s$, where the sound speed is
$c_s=150 \kms (T_1/10^6{\rm K})^{1/2}$.
For all the cases, we assume a constant cutoff momentum, $p^*=10^6 {\rm GeV}/c$, 
which is close to the knee energy in the Galactic cosmic ray spectrum.
For typical hot-phase ISM, $\delta =v_A/c_s \approx 1$ as mentioned before.
For typical ICM, $n_H \approx 10^{-3} {\rm cm}^{-3}$ and $B \approx 1-5 \mu$G, 
so $\delta \approx 0.5$ is taken here.
For typical test-particle limit solutions, we adopt $\epsilon_B=0.21$ 
to specify $p_{\rm inj}$ given in equation (\ref{pinj}), 
which determines the anchoring point where the test-particle power-law begins.
This choice of $\epsilon_B$ results in the injection rate $\xi \la 10^{-4}$
and the postshock CR pressure $P_{c,2}/(\rho_1u_s^2) \la 0.1$.
As can be seen in Figure 3, for stronger (faster) shocks, 
the postshock gas is hotter, the amplitude $f_{\rm inj}$ is higher and
the power-law spectrum is harder.

Then the CR pressure at the shock position can be calculated by
\begin{equation}
P_c(x_s)  = { {4\pi}\over 3} c \int_{p_{\rm inj}}^{\infty} f_{\rm tp}(x_s,p)
{p^4 dp \over \sqrt{p^2+(m_p c)^2}}.
\label{pcint}
\end{equation}
For strong shocks with $q_{\rm tp}=4$,  
with the test-particle spectrum in equation (\ref{ftest}),
$P_c \propto f_{\rm inj} p_{\rm inj}^4 \ln(p^*/m_p c)$.
Then, with a constant cutoff $p^*$, $P_c  \propto \exp(-Q_{\rm inj}^2) Q_{\rm inj}^4 p_{\rm th}$.
So for a fixed value of $Q_{\rm inj}$ (or fixed injection fraction $\xi$), 
$P_c \propto p_{\rm th} \propto u_s$. 
Figure 4 shows the fraction of injected particles and the postshock
CR pressure calculated by adopting the test-particle spectrum given in equation (\ref{ftest}).
The same $p^*=10^6 {\rm GeV}/c$ is chosen as in Figure 3.
The quantities, $n_{\rm cr,2}$ and $P_{c,2}$ do not depend sensitively on the assumed 
value of $p^*$ for weak shocks, since the power-slope $q_{\rm tp}$ is greater than 4.
But for strong shocks ($M\ga 30$) where $q_{\rm tp} \approx 4$ (see Fig. 1), 
the CR pressure increases logarithmically as $P_c \propto \ln (p^*/m_p c)$.
Several values of $T_1$, $\epsilon_B$ (or $Q_{\rm inj}$), and $\delta = v_A/c_s$ are considered. 
In general, for fixed values of $\epsilon_B$ (or $Q_{\rm inj}$) and $\delta$, the ratio
$P_{c,2}/(\rho_1u_s^2)$ increase strongly with the shock Mach number for shocks with $M\la 10$,
because of the strong dependence of $\xi$ (or $Q_{\rm inj}$) on $M$ for weaker shocks.
But for shocks with $M>10$, $\xi$ becomes independent of $M$ 
and so $P_c \propto u_s$, as discussed above. 
So the CR pressure relative to the shock ram pressure, $P_{c,2}/(\rho_1u_s^2) \propto u_s^{-1}$,
that is, it becomes smaller at faster shocks.
Of course, in the nonlinear DSA regime, the ratio $P_{c,2}/(\rho_1u_s^2)$ increases with
the shock Mach number and saturates at about 1/2 \citep{krj09}. 

The top panels of Figure 4 show how the CR pressure depends on $\epsilon_B$ and $\delta$.
For a given Mach number, the CR pressure increases strongly with 
$\epsilon_B$, because of the $\exp(-Q_{\rm inj}^2)$ factor.
Obviously, the CR pressure becomes smaller for larger $\delta$ 
because of softer power-law spectra at weaker shocks with $M\la 30$.
For $\epsilon_B=0.21$ and $\delta=1$, $\xi \la 10^{-4}$ and
$P_{c,2}/(\rho_1u_s^2) \la 0.1$,
so the test-particle solution would provide a good approximation. 
For $\epsilon_B=0.23$, on the other hand, the injection fraction becomes 
$\xi \approx 10^{-4}-10^{-3}$, and the test-particle solution is no 
longer valid for $M\ga 5$.
For weak cosmological shocks with $M\la 3$, 
typically found in the hot ICM \citep[\eg][]{ryuetal03,kangetal07}, 
even for a rather large value of $\epsilon_B=0.27$, 
the injection fraction is smaller than $10^{-3}$ and $P_{c,2}/\rho_1u_s^2 <0.01$ 
So we can safely adopt the test-particle solution for those weak shocks,
unless there are abundant pre-existing CRs in the preshock flow.

The middle panels show the cases with the same $Q_{\rm inj}$,
independent of $M$. For these cases, $T_1=10^6$K, $\delta=1$,
and $p_{\rm max}=10^6 {\rm GeV}/c$.
With the same $Q_{\rm inj}$, the injection fraction is almost independent 
of $M$ except for weak shocks with $M \la 5$.
For $Q_{\rm inj}=3.8$, $P_{c,2}/\rho_1u_s^2 \la 0.1$ for all shocks.
One can see that $Q_{\rm inj} \approx 3.8$ is the critical value, above which
the injection fraction becomes $\xi \la 10^{-4}$ and the ratio
$P_{c,2}/(\rho_1u_s^2) \la 0.1$.  
Hence, if $p_{\rm inj} \ga 3.8 p_{\rm th}$, the CR injection fraction is small enough to
guarantee the validity of test-particle solution.
But once again one should note that $P_c \propto \ln p^*$ for strong shocks.

The bottom panels show the cases in which the preshock temperature is $T_1=10^5-10^7$K.
Since the ratio $P_{c,2}/(\rho_1u_s^2) \propto \xi u_s^{-1}$ and $\xi$ does not
depend on $T_1$, $P_{c,2}/(\rho_1u_s^2) \propto \xi T_1^{-1/2}$
for a given Mach number, $M=u_s/c_s$.  
So we chose $\epsilon_B \approx 0.20-0.22$ for different $T_1$,
which results in $\xi \sim 10^{-4}(T_1/10^6{\rm K})^{1/2}$.
This gives the similar value of $P_{c,2}/(\rho_1u_s^2) \sim 0.1$ 
for three values of $T_1$. 
For these shocks, the test-particle solution would be valid.

When $P_{c,2}/(\rho_1u_s^2) >0.1$, 
the nonlinear feedback of the diffusive CR pressure becomes important
and the evolution of CR modified shocks should be followed by DSA simulations.
Figure 5 compares the evolution of a slightly modified $M=5$ 
shock ($\epsilon_B=0.27$) with that of a test-particle shock ($\epsilon_B=0.2$).
In the CR modified shock, the upstream flow is decelerated in the precursor before it enters
the gas subshock. So the quantities at far upstream, immediately upstream and downstream 
of the subshock are subscripted with '0', '1', and '2', respectively.
For the test-particle shock, $\rho_1=\rho_0$ and $T_1=T_0$.
Here $T_0=10^6$K and $v_A/c_s= 0.42$.
The simulations start with a purely gasdynamic shock 
at rest at $x = 0$, initialized according to Rankine-Hugoniot relations 
with $u_0 = -1$, $\rho_0 = 1$ and a gas adiabatic index, $\gamma_g = 5/3$.
There are no pre-existing CRs.

The test-particle spectrum given in equation (\ref{ftest}) 
with $p^*=1.2 p_{\rm max}$ at $t/t_0=10$ is also shown for comparison (dot-dashed lines)
in the bottom panels.
In the test-particle shock with $\epsilon_B=0.2$, both $P_{c,2}/(\rho_0u_s^2)\approx 0.005$ 
and $f(x_s)$ from the DSA simulation agree well with the test-particle solution given in 
equation (\ref{ftest}), as expected.

If we were to take the test-particle spectrum with $\epsilon_B=0.27$, we would obtain
$\xi = 1.74 \times 10^{-3}$ and $P_{c,2}/(\rho_1u_s^2) = 1.17$,
which is unphysical.
In the CR modified solution from the DSA simulation, however, 
$\xi \approx 3.6 \times 10^{-4}$ and $P_{c,2}/(\rho_1u_s^2) \approx 0.1$. 
The postshock temperature $T_2$ is reduced about 17 \% in the CR modified solution 
(due to higher $\rho_2$ and lower $p_{g,2}$), 
compared to that in the test particle solution. 
But $u_2$ and so $p_{\rm inj}$ remain about the same.
As a result, the amplitude $f_{\rm inj}$ is lower than that of the test-particle spectrum (see the bottom right panel of Fig. 5)
and so the injection rate is reduced in the CR modified solution.
The distribution function $f(x_s,p)$ from the DSA simulation is slightly 
steeper for $p/m_p c<10$ and slightly flatter
for $p/m_p c>10$ than the test-particle power-law, 
because the flow velocity is slightly modified.
This demonstrates that the DSA saturates in the limit of efficient injection 
through the modification of the shock structure (\ie a precursor plus a weak gas subshock), 
which in turn reduces the injection rate. 
Thus the ratio $P_{c,2}/(\rho_1u_s^2)$ approaches to $\sim 1/2$ for strongly modified CR shocks
\citep{kj07}.

Finally, we find that the volume integrated spectrum contained in the simulation volume 
can be obtained simply from $F(p)=\int f(x,p) dx \approx f_{\rm tp} (x_s,p) u_2 t$.
This provides the total CR spectrum accelerated by the age $t$. 

\section{SUMMARY}

Although the nonlinear diffusive shock acceleration (DSA) involves
rather complex plasma and MHD processes, the test-particle solution
may unveil some simple yet essential pictures of the theory.
In this study, we suggest an analytic form for the CR spectrum
from DSA in the test-particle regime,
based on simple models for thermal leakage injection and 
Alfv\'enic drift of self-generated resonant waves.

If the particle diffusion is specified (\eg Bohm diffusion),
the shock Mach number is the primary parameter
that determines the efficiency of diffusive shock acceleration.
For a given shock Mach number,
the fraction of injected CR particles becomes the next key factor. 
Since the postshock thermal velocity distribution
at the injection momentum determines the amplitude of the 
power-law distribution in the thermal leakage injection model, 
the ratio $Q_{\rm inj} = p_{\rm inj}/p_{\rm th}$ is the key parameter
that controls the CR injection fraction and in turn determines the CR
acceleration efficiency. 
On the other hand, as a result of the drift of Alfv\'en waves in the
precursor, the power-law slope should be revised as in equation (\ref{qtp}),
which leads to the CR spectrum much steeper than the canonical test-particle
power-law. This effect is negligible for shocks with the Alfv\'enic Mach 
number, $M_A \ga 30$.

For shocks with the sonic Mach number $M\ga 10$,
depending on the preshock temperature $T_1$, 
the injection fraction, $\xi \la \xi_c \approx 10^{-4}(T_1/10^6{\rm K})^{1/2}$ 
would lead to the downstream CR pressure, $P_{c,2}/(\rho_1u_s^2)\la 0.1$.
The exact values depend on other parameters such as $v_A$.
In that case, the CR spectrum at the shock location can be
described by the test-particle power-law given in equation (\ref{ftest}),
in which the amplitude, $f_{\rm inj}$, is fixed by the postshock
thermal distribution at $p_{\rm inj}$ given in equation (\ref{finj}). 
For supernova remnants in the hot-phase of the ISM with $T_1=10^6$K, for example,
the CR injection fraction becomes less than $10^{-4}$,
if $Q_{\rm inj} \ga 3.8$ (or $\epsilon_B\la 0.21$). 
For weaker shocks with $M<5$, the test-particle solution is valid even for
larger injection fraction, so $\xi_c<10^{-3}$.

We have shown that the CR spectrum at the shock location in time-dependent,
test-particle shocks without particle escape could be approximated by 
the analytic solution given in equation (\ref{ftest}),
which was derived for steady-state, test-particle shocks by \citet{capri09}, 
with the cutoff momentum set as $p^* \approx 1.2 p_{\rm max}(t)$.
If the CR injection is inefficient, which should be true 
for weak shocks with $M\la 5$ found in the intracluster medium, 
the test-particle solution presented in this paper should 
provide a good approximation.
Figure 4 should provide guidance to assess if a shock with specific
properties can be treated with the test-particle solution. 

With the injection rate greater than $\xi_c$,
especially for shocks with $M>5$, the spectrum deviates from the
test-particle form due to the modified flow structure caused by
the diffusive CR pressure.
In fact, the DSA efficiency saturates in strongly modified CR shocks,
because the postshock temperature gets lower and so the injection rate is reduced. 
Based on the results of the DSA simulations, \citet{krj09} suggested that
CR-modified shocks evolve self-similarly
once the total pressure is dominated by relativistic particles, 
and that the CR spectrum at the subshock can be approximated by
the sum of two power laws with the slopes determined by the subshock and 
total compression ratios with an exponential cutoff at 
$p_{\rm max}(t)$.

\acknowledgements
The authors would like to thank T. W. Jones for helpful comments on the paper.
HK was supported by National Research Foundation of Korea
through grant 2009-0075060.
DR was supported by National Research Foundation of Korea through grant
KRF-2007-341-C00020.

\clearpage

\begin{figure}
\begin{center}
\includegraphics*[height=40pc]{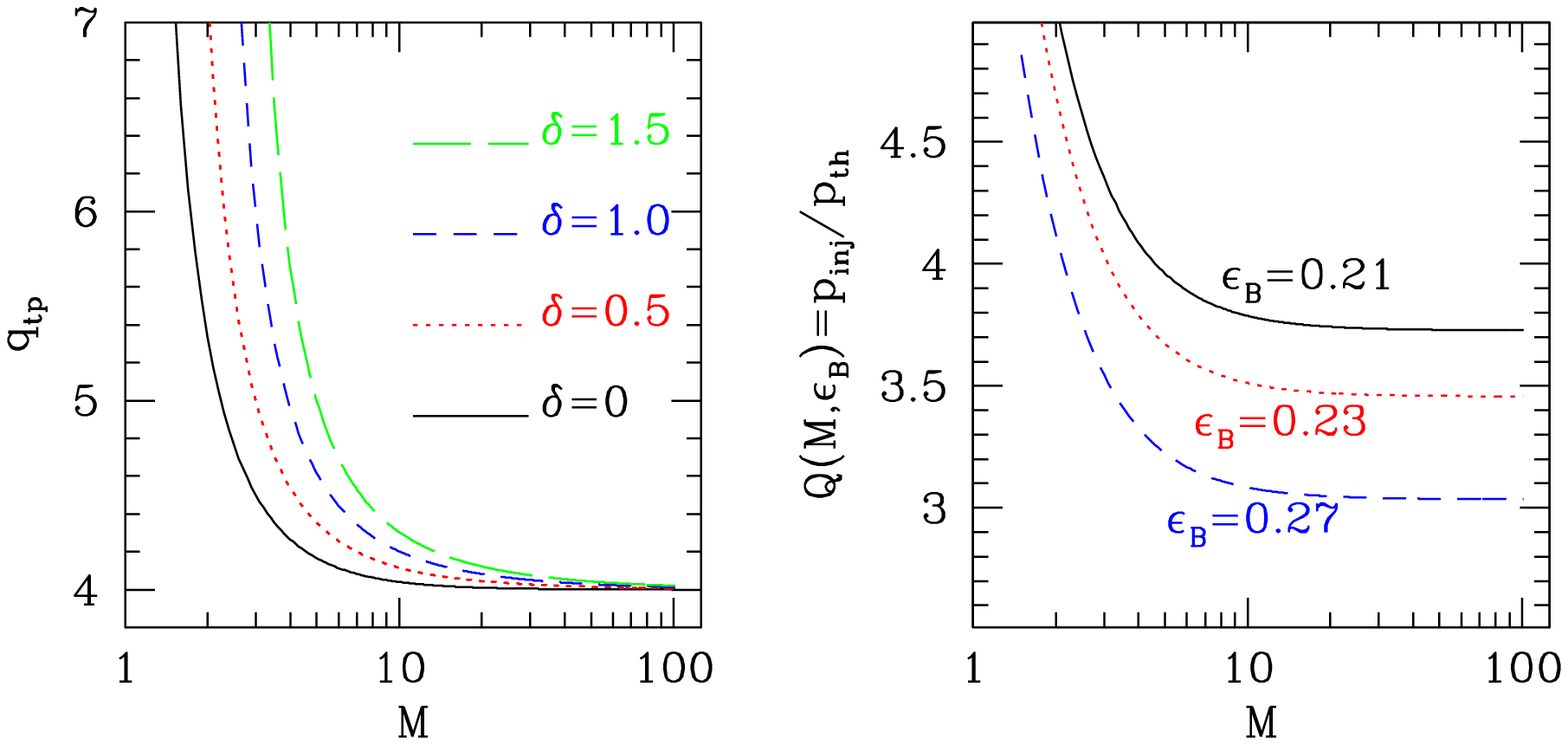}
\end{center}
\vspace{-8.0cm}
\caption{
{\it Left:}
The test-particle power-law slope, $q_{\rm tp}$,
revised by including the Alfv\'enic drift (Eq. [\ref{qtp}]), is shown as a function of sonic Mach number
for four different values of $\delta= v_A/c_s$.
{\it Right}: The ratio $Q_{\rm inj}=p_{\rm inj}/p_{\rm th}$ is shown for three values of
$\epsilon_B= B_0/B_{\perp}= 0.21$, 0.23 and 0.27.
}
\end{figure}
\clearpage

\begin{figure}
\begin{center}
\includegraphics*[height=40pc]{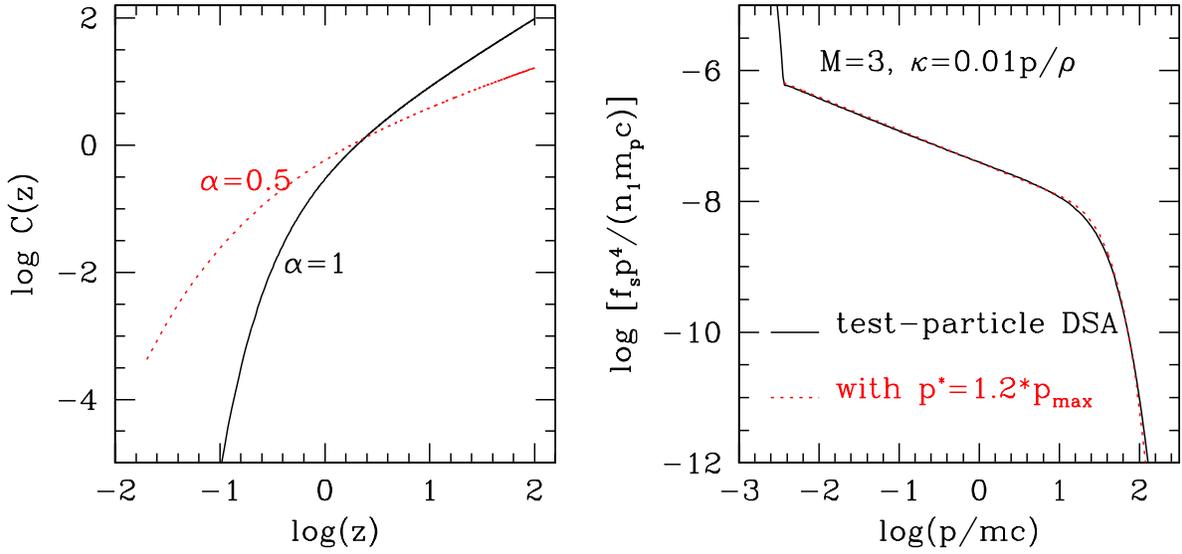}
\end{center}
\vspace{-8.0cm}
\caption{
{\it Left:} The function $C(z)$ defined in Eq. (\ref{cofz}) is shown 
as a function of $z\equiv p/p^*$ 
for $\kappa(p)=\kappa^* (p/m_p c)^{\alpha}$.
{\it Right:} The CR distribution at the shock position, $f_s \cdot p^4$ 
(in units of $n_1 m_p c$), is shown for a Mach 3 shock.
The solid line shows the results of a time-dependent DSA simulation
without particle escape in the test-particle regime (\ie no CR feedback to the flow). 
The dotted line represent
the test-particle spectrum given in Eq. (\ref{ftest}) with $p^*= 1.2
p_{\rm max}$.  }
\end{figure}
\clearpage

\begin{figure}
\begin{center}
\includegraphics*[height=40pc]{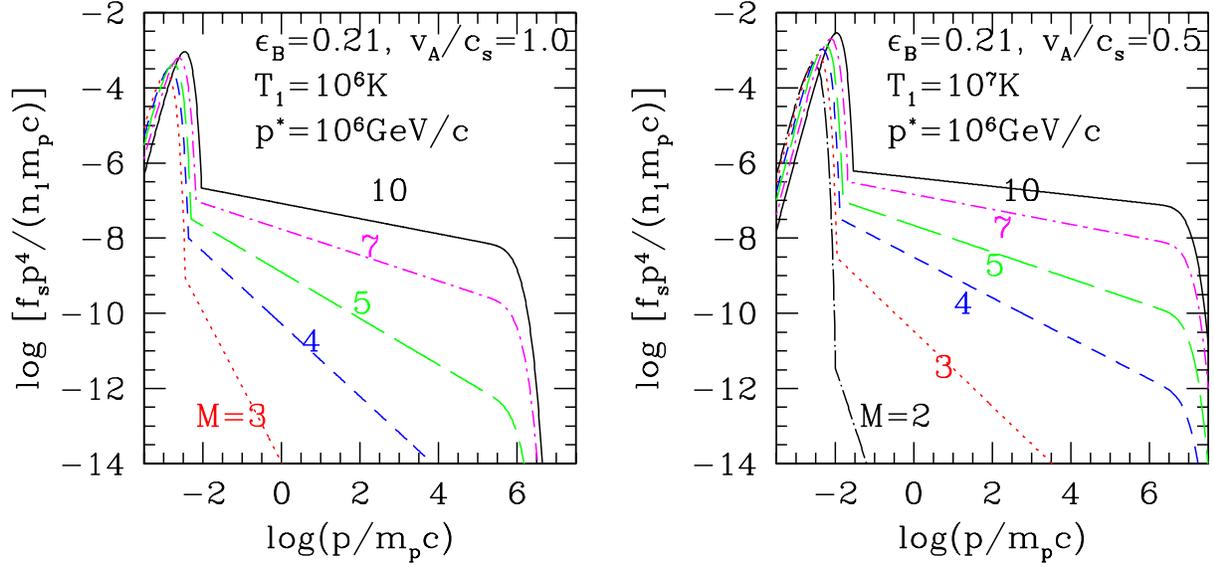}
\vspace{-8.0cm}
\end{center}
\caption{
{\it Left}: Test-particle spectra given in Eq. (\ref{ftest}) with
$\epsilon_B=0.21$, $p^*=10^6{\rm GeV}/c$, $T_1=10^6$K, and $\delta =v_A/c_s= 1.0$.
{\it Right}: Same as the left panel except $T_1=10^7$K, and $v_A/c_s= 0.5$. 
Each curve is labeled with the shock Mach number, $M$, and the
shock speed is $u_s= M\cdot 150 \kms (T_1/10^6{\rm K})^{1/2}$.
}
\end{figure}
\clearpage

\begin{figure}
\begin{center}
\includegraphics*[height=40pc]{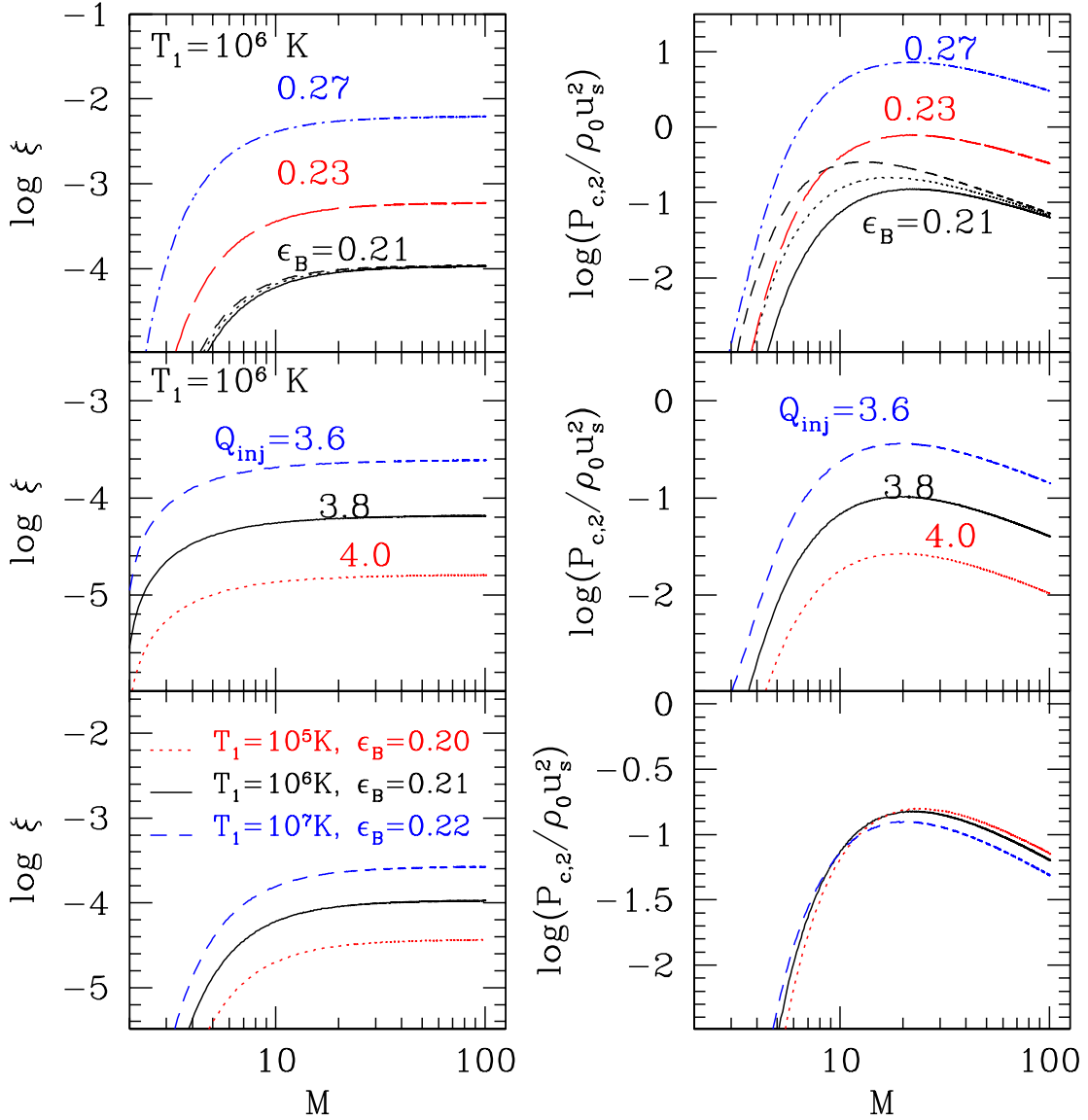}
\vspace{-2.0cm}
\end{center}
\caption{
The fraction of CR particles, $\xi = n_{\rm cr,2}/n_2$ (left panels), 
and the downstream CR pressure in units of the shock ram pressure (right panels)
are shown  for the test-particle spectrum given in Eq. (\ref{ftest})
with a fixed $p^*=10^6 {\rm GeV}/c$. 
The shock speed is specified by $u_s=M\cdot 150 \kms (T_1/10^6{\rm K})^{1/2}$.
{\it Upper panels}:
Shocks with the preshock temperature, $T_1=10^6$ K.
Three values of $\epsilon_B=0.21$,  0.23 (long dashed lines), 
and 0.27 (dot-dashed) are considered with $\delta = v_A/c_s= 1.0$.
For the case with $\epsilon_B=0.21$, three cases with
$\delta= 0$ (dashed lines), 0.5 (dotted), and 1.0 (solid) are shown.
{\it Middle panels}:
The same model shocks as the upper panels except a constant ratio
$Q_{\rm inj}=p_{\rm inj}/p_{\rm th}= 3.6$ (dashed lines), 3.8 (solid), 4.0 (dotted),
are shown. The Alf\'en speed is $v_A/c_s= 1.0$. 
{\it Lower panels}:
Shocks propagating into different temperature media,
$T_1=10^5$K (with $\epsilon_B=0.20$, dotted lines), 
$10^6$K (with $\epsilon_B=0.21$, solid lines), 
and $10^7$K (with $\epsilon_B=0.22$, dashed lines) are shown.
The Alf\'en speed is $v_A/c_s= 1.0$ for all three cases. 
}
\end{figure}
\clearpage

\begin{figure}
\begin{center}
\includegraphics*[height=40pc]{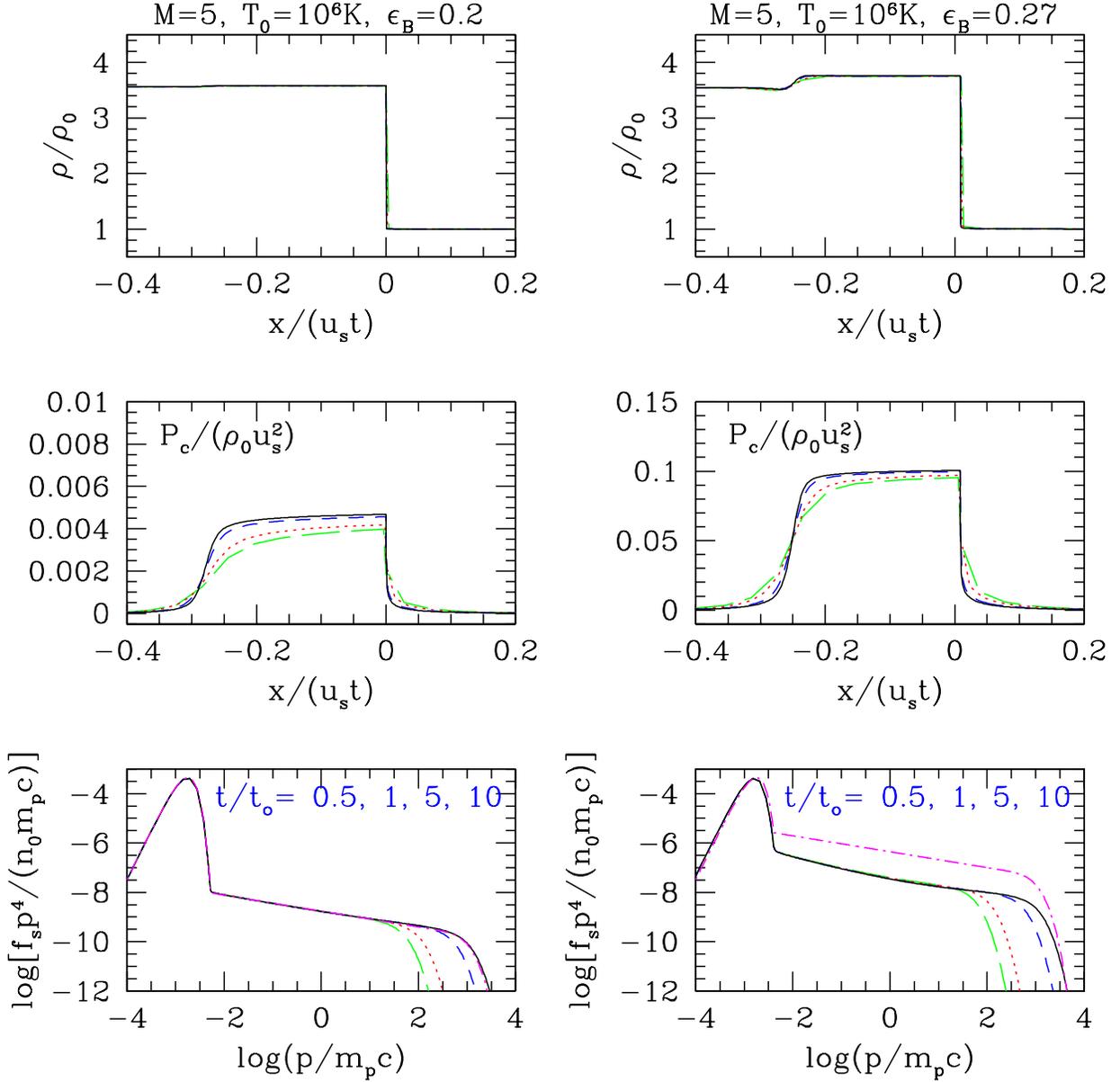}
\end{center}
\caption{
Time-dependent DSA simulation results for a $M=5$ shock for $\epsilon_B=0.20$
(low injection rate, left panels) and $\epsilon_B=0.27$
(high injection rate, right panels). 
Here $T_1=10^6$K, $v_A/c_s= 0.42$ and $\kappa=10^{-3}(p/m_p c)(\rho_0/\rho)$.
At the last time epoch ($t/t_o=10$) the cutoff momentum becomes
$p^* =1.2 p_{\rm max}\approx 10^3{\rm GeV}/c$.
The shock structure is shown at $t/t_o=$ 0.5 (long dashed lines),
1 (dotted), 5 (dashed), and 10 (solid) as a function
of the similarity variable, $x/(u_s t)$.
The (magenta) dot-dashed lines in the bottom panels
represent the test-particle spectra given in Eq. (\ref{ftest})
at $t/t_o=10$.
Note that the dot-dashed line (analytic solution) 
and solid line (numerical solution) almost coincide with each other
in the case with $\epsilon_B=0.2$. 
Here $t_o$ is a normalization constant.
}
\end{figure}

\end{document}